# Title: Subsoil acidity causes long delays in inorganic carbon sequestration by Enhanced Weathering

**Short title: Subsoil acidity may undermine Enhanced Weathering**

Søren Jessen[a,*], Rasmus Jakobsen[b], Majken Looms[a], Per Ambus[a] and Dieke Postma[b]

[a]Department of Geosciences and Natural Resource Management, University of Copenhagen, Øster Voldgade 10, 1350 Copenhagen K, Denmark

[b]Department of Geochemistry, Geological Survey of Denmark and Greenland, Øster Voldgade 10, 1350 Copenhagen K, Denmark

*sj@ign.ku.dk

## Abstract

While a looming atmospheric $CO_2$ overshoot calls for immediate carbon sequestration, delays associated to Enhanced Weathering (EW) carbon dioxide removal are being investigated. Topsoil acidity is already known to delay EW carbon sequestration, but subsoil acidity remains underexplored. Using century-long agricultural liming of formerly acidic heathland as a proxy for EW, this study provides empirical evidence of subsoil-imposed delays. Below such limed terrain, we observed a downward-progressing front of topsoil-produced alkalinity that still requires 30–100 years to penetrate the ~5 m thick acidic sandy unsaturated zone and reach the groundwater table. Subsoil acidity thus may cause beyond-reasonable delays, prohibiting EW as a viable short-term carbon capture strategy even on topsoils made non-acidic by preceding liming. When planning EW schemes, the amounts of stored acidic cations in top- and subsoil, as well as the rate and composition of infiltrating water, controlling the duration of the delay, require careful assessment.

### Teaser

The study of a century of agricultural liming indicate that Enhanced Weathering as a soil treatment may fail to remove atmospheric $CO_2$ in time.

## Introduction

It is becoming increasingly clear that the targets of the Paris climate agreement will be missed and that the Earth will face a global warming overshoot scenario (*1*). Limiting global warming therefore requires immediate $CO_2$ removal from the atmosphere, until emissions are brought under control.

Research currently investigates the potential of speeding up silicate weathering via Enhanced Weathering (*2*–*7*) (EW) as a negative emission technology. EW harnesses reactions between $CO_2$ and silicate minerals that convert soil gas $CO_2$—ultimately derived from the atmosphere (*8*)—into carbonate alkalinity (*9*) or carbonate minerals (*5*, *10*). Proposed EW schemes make use of existing infrastructure and technology to distribute fine-grained silicates to agricultural soils and further benefits from the increased weathering reaction rates at the elevated partial pressure of $CO_2$ in soils relative to the atmosphere (*11*, *12*). Global sequestration rates possible by EW have been assessed to be in the Gt $CO_2$ $yr^{-1}$ range (*2*, *13*), indicating very significant potentials.

The $CO_2$ sequestration by EW is mainly ascribed to the increase in the advective downward flux of carbonate alkalinity (i.e., dissolved bicarbonate and carbonate ions, henceforth referred to as alkalinity), formed by the reaction of carbonic acid with silicate minerals added as amendment products to the topsoil (i.e., the plow layer), and ultimately its export to the ocean (*14*). Particular EW interests are associated to areas in need for soil acidity mitigation (*4*), because such areas contain both the potential for an enhanced alkalinity production at enhanced feedstock dissolution rates, and the benefits of enhanced crop production resulting from the concurrently increased soil pH and co-released nutrients (*3*, *11*). With few exceptions (*15–18*), field assessments of the alkalinity production by EW are based on laboratory derived reaction kinetics of mineral dissolution or gas exchange, where the rates are enhanced by increased temperatures, low pH and small grain size, and the water residence time in the soil zone (*2*) allowing for the weathering reactions to proceed. Furthermore, EW alkalinity production rates are frequently assessed with observations made in the topsoil (*7*, *19*), such as via measurement of electrical conductivity in topsoil pore water (*20*), or observations of feedstock dissolution (*21–23*). Unfortunately, such assessments of EW alkalinity production do not account for the fate of the alkalinity following its production in the topsoil.

For short-term carbon dioxide removal to occur, water containing alkalinity must at least reach the saturated groundwater zone. However, recent research have suggested years to decades delays between the alkalinity being formed by dissolution of topsoil EW amendment products and the onset of carbon sequestration (*14*, *24*). Cation exchange in acidic topsoils has been of particular concern (*5–7*, *24*). For example, Kanzaki et al. (*24*) outlines how the exchange of feedstock base cations for acidic cations may lead to transformation of bicarbonate back into $CO_2$, thus cancelling the soil-zone $CO_2$ sequestration. Therefore, studies accounting for topsoil acidity have suggested that soil acidity compromises EW efficiency, by delaying $CO_2$ sequestration, despite promoting mineral dissolution (*14*, *25*). Alike geochemical processes have been suggested to delay the passage of alkalinity across deeper strata (*6*), although empirical evidence is lacking. Subsoil acidity may particularly easily be overlooked in already-limed soils that lack topsoil acidity and thus appear suitable for EW deployment. Such soils often occur in developed regions where EW deployment is otherwise facilitated by existing infrastructure and ease of implementing economic incentives.

Here, we use agricultural liming as a roughly century-long proxy experiment for an EW scheme to study the fate of topsoil-generated alkalinity as it moves downward through the unsaturated zone and into the underlying aquifer.

## Results and Discussion

To trace the fate of topsoil-produced alkalinity, we sampled pore gas and pore water through a ca. 5 m thick unsaturated zone of an agricultural field on post-glacially weathered Quaternary glacial outwash sands (*26*) (Supplementary Materials, Fig. S1). Historical maps show that the field site was part of western Denmark's extensive heathland until its reclamation between 1874 and 1897 (Fig. S2). Further, records from national campaigns to distribute marl and lime to Danish acidic soils indicate that the field site received ~10 $m^3$/ha of marl between 1906 and 1910, and lime from the 1930'ies (Figs. S1 and S3). Accordingly, prior to our sampling, the field site has been subject to agricultural liming for close to a century. The annual precipitation surplus is around 500 mm (*27*) and the volumetric water content was found to be stable near 0.10 throughout the year. These figures correspond to an estimated average residence time of water in the unsaturated zone of ca. 1 yr. The stable isotopes composition of the water at each sampling depth varied annually, indicating that the soil water was not stagnant (Fig. S4). In support, core materials (*26*) and the uniform soil gas $CO_2$ profiles (see below) indicated absence of subsoil low permeability zones.

Parameters of the inorganic carbonate system are shown in Fig. 1; for clarity, only profile data from one out of four experimental plots are included. For data from all plots see Supplementary Materials. Fig. 1A shows that pore gas $P_{CO_2}$ was highest in the autumn at 0.06 atm and then decreased to minima at ~0.02 atm during winter until the onset of spring respiration in the following growth season. Peak $P_{CO_2}$ during spring and summer occurred at ca. 0.5 m depth. The $P_{CO_2}$ profile dynamics are consistent with a diffusive downward $CO_2$ gas transport to the deeper unsaturated zone during spring and summer, followed by 'degassing' to the atmosphere by upwards diffusive transport during winter (*28*). Given a porosity of 0.3, the amount of carbon present as gaseous $CO_2$ in the ~5 m thick unsaturated zone varies seasonally between roughly 880 and 2550 mmol C $m^{-2}$. The unsaturated zone remained oxic, and $P_{CO_2}$ values were always near the deficit in $P_{O_2}$ from the atmospheric $P_{O_2}$ of 0.21 atm, as predicted by oxic respiration control on $CO_2$ and $O_2$ partial pressures. The higher unsaturated zone $P_{CO_2}$ in autumn rather than summer may be explained by the concurrent root degradation and moist topsoil with lowered gas diffusivities in the wetter autumn.

Before the onset of agricultural liming, the Quaternary outwash plain's heathland soil and subsoils were naturally acidic (e.g., pH ~5 (*29–31*)). The pore water under such conditions would possess only diminutive or negative alkalinities. However, the upper 2m of the current profile contained pore water with significant carbonate alkalinities (Fig. 1B; shaded areas), along with a 1-to-2-unit higher pH value (Fig. 1C), and a Ca-dominated cation exchanger (Fig. 1E). Below 2 m depth, we observed a marked drop in alkalinity and pH, along with a shift from a Ca- to an Al-dominated cation exchanger. We interpret the higher pH and exchanger Ca-dominance in the upper 2 m to be a result of agricultural liming. Vise-versa, we interpret the low pH and exchanger Al-dominance below 2 m as reflecting conditions before the onset of agricultural liming.

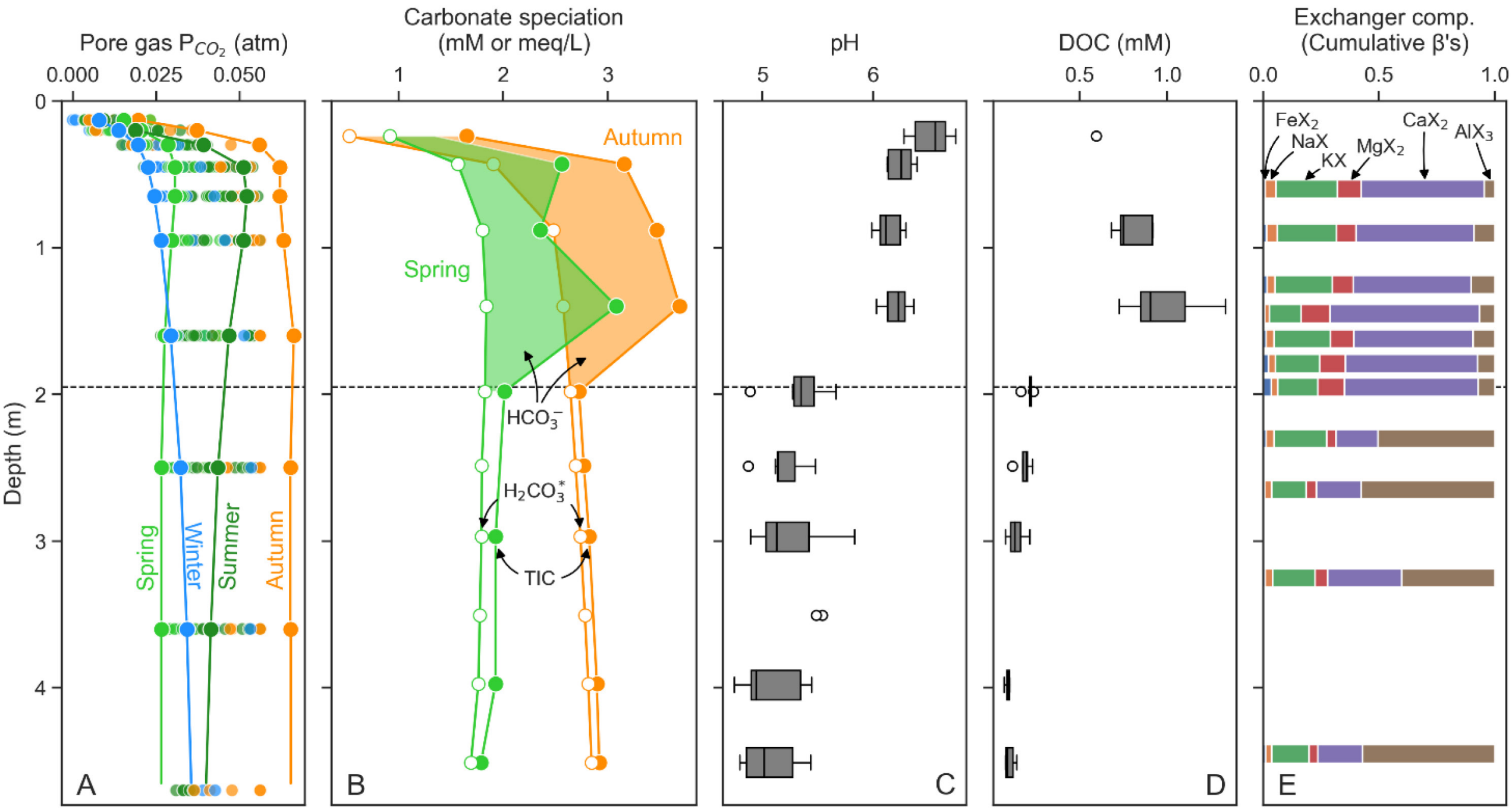

**Fig. 1. Summary of observations from one of four profiles at the Enhanced Weathering test field. (A) Pore gas $P_{CO_2}$. (B) Inorganic carbonate speciation of pore water in spring (lowest $P_{CO_2}$) and autumn (highest $P_{CO_2}$). (C) Pore water pH. (D) Dissolved organic carbon (DOC). (E) Cation exchanger composition given as equivalent fractions. Boxes in box diagrams are based on $n$ = 6–10 (C) and $n$ = 4–5 (D) measurements. The total number of measurements were $n$ = 84 for pH and $n$ = 35 for DOC. Open circles in box diagrams are statistical outliers to the corresponding box, or data points when $n$ < 3 and no box is drawn. The groundwater table is ca. 5 m below surface. The dashed line shows depth of the alkalinity front (pH front) as interpreted from combined data in panels B, C, D and E.**

The amount of carbonic acid (defined as $H_2CO_3^*$ = sum of $CO_{2(aq)}$ and $H_2CO_3$) varied seasonally between 2 mM (spring) and 3 mM (autumn) as controlled by equilibrium with the soil gas $P_{CO_2}$ (*28*) (Fig. 1B):

$$CO_{2(g)} + H_2O \leftrightarrow H_2CO_3^* \qquad \text{(Reaction 1)}$$

This corresponds to a seasonal variation in the amount of carbon stored as dissolved carbonic acid ($H_2CO_3^*$) between roughly 1000 (spring) and 1500 (autumn) mmol C m$^{-2}$. Adding dissolved carbonic acid to the above-calculated carbon stored as $CO_2$ in the air-filled pore space yields a total variation between 1880 and 4050 mmol C $m^{-2}$. The cyclic twofold increase and decrease in the sum of gaseous and dissolved $CO_2$ translates to an average residence time of $CO_2$ of roughly 2 yr in the unsaturated zone.

The results from a summer and a winter sampling campaign in Fig. 1B showed a seasonally invariable penetration depth of alkalinity of 2 m, indicating the removal of alkalinity from the pore water as the percolating water passes this depth. The seasonal variation in temperature within the unsaturated zone was between 1°C and 21°C at 5 cm depth and 4°C and 14°C at 2 m depth. Given that abiotic processes have been observed to be less temperature dependent in comparison to biotic processes (*32*, *33*), the lack of temperature dependence for the alkalinity penetration depth indicate that abiotic (inorganic), rather than biotic processes, control the loss of alkalinity from solution at 2 m depth.

Carbonate dissolution liberates fossil carbon and therefore is inadequate as EW feedstock. Nevertheless, carbonate dissolution by carbonic acid releases one base cation charge for each generated alkalinity charge, just like other soil amendment products proposed for EW, such as olivine-rich ultramafic and mafic rocks, basaltic rocks, or concrete waste (*34*). Reaction 2 illustrates the dissolution reaction in an alkaline topsoil, using calcite as an example:

$$CaCO_3 + H_2CO_3^* \leftrightarrow Ca^{2+} + 2HCO_3^- \qquad \text{(Reaction 2)}$$

This makes agricultural liming a useful proxy to study the long-term fate of the topsoil-produced alkalinity of terrestrial EW schemes. Depending on the EW amendment product (excepting carbonates) and the hydrological setting, the objective of EW is for the produced bicarbonate ions to increase groundwater alkalinity, or precipitate as stable pedogenic carbonates further down the flow path, or ultimately reach the ocean (*5*, *6*), in all cases effectively removing $CO_2$ from the atmosphere. In our experimental plots, though, pore waters were always subsaturated for calcite dismissing pedogenic carbonate precipitation. Therefore, other processes must be responsible for the loss of alkalinity in Fig. 1B.

It is worth noting the substantial amount of alkalinity that is lost: given the precipitation surplus of ~500 mm $yr^{-1}$ and a loss of ~1 meq alkalinity per liter, the rate of the alkalinity loss amounts to ~500 meq $m^{-2}$ $yr^{-1}$. Such a high rate substantiates the involvement of major chemical components rather than of minor trace components. Disregarding the presence of nitrate, the generally dominant dissolved ions above the alkalinity front were Ca, Mg and bicarbonate. The Ca and Mg ions present in the acidic water, below the alkalinity front, indicate the leaching of divalent base cations, balanced by other anions than bicarbonate or carbonate. While some literature proposes carbon sequestration via ocean export of base cations balanced by other anions than aqueous carbonates (*6*, *14*, *35*), the present paper focuses on explaining the loss of carbonate alkalinity observed at 2 m depth (Fig. 1B), as this alkalinity is a direct carrier of carbon.

**Loss of alkalinity by cation exchange and mineral equilibrium**

The loss of alkalinity is explained by a reaction sequence in which Ca released by lime dissolution (e.g., reaction 2) displaces exchanger Al to the pore water, followed by Al hydroxide precipitation (*24*, *34*, *36*, *37*). The formation of $Al(OH)_3$ releases three protons per Al precipitated. The combined reaction, with $Ca^{2+}$ as example, therefore becomes:

$$Ca^{2+} + \tfrac{2}{3}AlX_3 + 2H_2O \leftrightarrow CaX_2 + \tfrac{2}{3}Al(OH)_3 + 2H^+ \qquad \text{(Reaction 3)}$$

in which X is a negative exchanger site. Following reaction 3, the alkalinity co-produced by lime dissolution becomes consumed by the fast reaction between $H^+$ and bicarbonate:

$$H^+ + HCO_3^- \leftrightarrow H_2CO_3^* \qquad \text{(Reaction 4)}$$

The carbon in the carbonic acid produced by reaction 4 may then be lost as $CO_2$ to the soil gas by the reversed reaction 1, and subsequently returned to the atmosphere by upwards gaseous diffusion.

The $CO_2$ generation by alkalinity neutralization might cause one to suspect a peak in soil gas $P_{CO_2}$ at the depth of the front. At the site, we measured an effective gaseous diffusion coefficient $D_e$ of 1.2–6×10$^{-6}$ m$^2$ s$^{-1}$ in the unsaturated zone (*38*, *39*). Using Fick's first law, and assuming a steady-state $P_{CO_2}$ at 0.5 m depth and no $CO_2$ production below 0.5 m except for the $CO_2$ release at 2 m depth at a rate of 500 mmol m$^{-2}$ yr$^{-1}$ would increase $P_{CO_2}$ by only up to 0.46×10$^{-3}$ atm at and below 2 m depth (Supplementary Text 2 and Fig. S6). This is too small to be observable relative to the observed soil gas $P_{CO_2}$ range of 20×10$^{-3}$ to 60×10$^{-3}$ atm (Fig. 1A).

Apart from the loss of dissolved inorganic carbon, a loss of dissolved organic carbon (DOC, Fig. 1D) was observed concurrent with the drop in pH at 2 m depth (Fig. 1C). NaOH-extracts of sediments in the profile of Fig. 1 showed that a peak of up to 20 mmol DOC kg$^{-1}$ dry sediment had accumulated at 2 m depth (Fig. S5), corresponding to the position of the shift in pH in Fig. 1 (dashed line). Because Al-hydroxide precipitation is known to effectively scavenge DOC (*40*), the DOC and NaOH-extraction profiles together support the position of an Al displacement front at 2 m depth.

In the idealized case where the only reactions occurring are *i*) calcite's dissolution by carbonic acid, followed by *ii*) Ca-for-Al cation exchange, *iii*) Al-hydroxide precipitation, *iv*) alkalinity neutralization and *v*) $CO_{2(g)}$ release, the net reaction is the sum of reactions 2, 3, two times 4, and one time reaction 1 reversed:

$$CaCO_3 + \tfrac{2}{3}AlX_3 + H_2O \leftrightarrow CaX_2 + \tfrac{2}{3}Al(OH)_3 + CO_{2(g)} \qquad \text{(Reaction 5)}$$

Accordingly, for each mole of calcite dissolved, the net result apparent from reaction 5 is the production of one mole of $CO_{2(g)}$ which stems from the carbonate ion in the calcite. Similar reaction chains can be derived for other minerals proposed for EW (e.g., forsterite, olivine, albite, anorthite) in the presence of $CO_2$ under the assumption that their released cation(s) will displace Al from an acidic exchanger. Such reaction chains produce similar results in that the production of alkalinity in solution balances the production of cationic charges, which in turn balances the $H^+$ production stemming from cation exchange and Al-hydroxide precipitation.

In the above reaction scheme, the delay before carbon sequestration by EW begins is directly linked to the cation exchange capacity (CEC). Depending on the methodology, we observed top- and subsoil CECs of 5.3 and 0.56 meq/100 g, respectively, when determined as the sum of charges of individual cations displaced by $NH_4Cl$, and top- and subsoil CECs of 11.5 and 1.6 meq/100 g, respectively, when determined by Na displacement by $NH_4Cl$ after loading the exchanger complex with Na.

**Present-day downward progression of the alkalinity front**

The current downward velocity of the alkalinity front ($v_{\text{front}}$) can be estimated from the mass transfer between the dissolved and exchanger-bound Ca (and Mg) in accordance with the above reaction scheme as water passes the alkalinity front, using the formula:

$$v_{\text{front}} = v_{H_2O}/R_{\text{front}} = v_{H_2O}/(1 + dq/dc) \qquad \text{(Equation 1)}$$

where $v_{H_2O}$ is the average pore water velocity (5 m yr$^{-1}$), $R_{\text{front}}$ is the retardation coefficient, and $dq$ and $dc$ are the changes in, respectively, exchanger-bound and solute Ca and Mg concentrations, all in units of mmol per L of contacting pore water. This manual estimation leverages the observed solute concentrations, Ca and Mg equivalent fractions, and avg. subsoil CEC span of 0.56 to 1.6 meq/100 g to yield a front velocity range of 3.0 to 8.7 cm yr$^{-1}$ (Supplementary Text 1). By extrapolation, it took the alkalinity front 20–57 yr to cover the 1.7 m vertical distance from the bottom of the topsoil at 0.3 m to 2 m depth, and another 35 to 100 yr will pass before the alkalinity front reaches the groundwater table at 5 m depth. Together, subsoil acidity accordingly delays $CO_2$ transfer to the aquifer by 55–157 yr. During this delay, no downward transport of alkalinity into the underlying aquifer will take place as the $CO_2$ produced at the alkalinity front will diffuse upwards to the atmosphere. If an EW scheme was to start today at the field site, more than three decades, and perhaps a whole century, would pass before direct $CO_2$ sequestration, related to topsoil alkalinity generation, would begin.

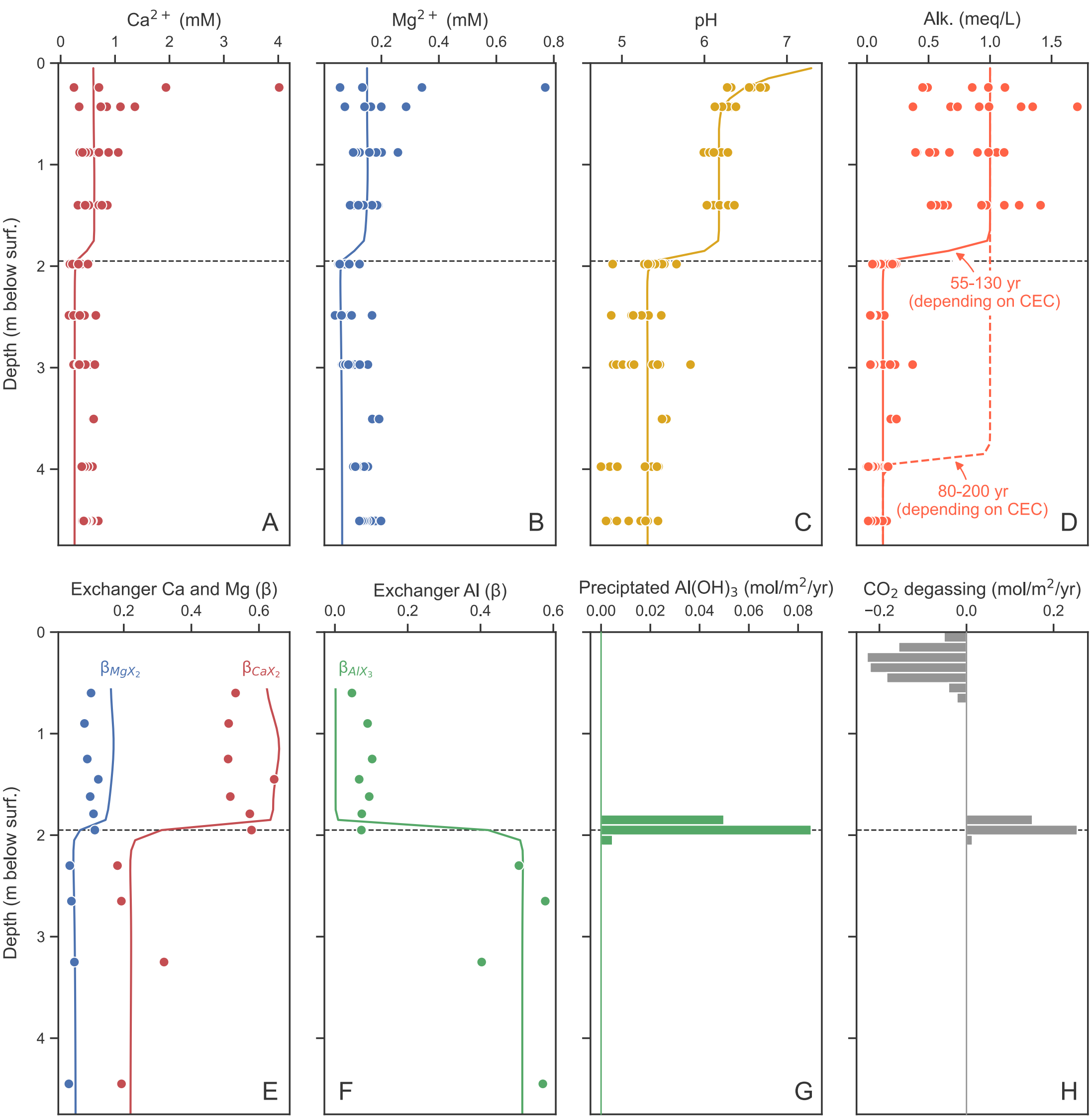

**Fig. 2. (A–F) Comparison of model simulation after 55–130 years of liming (full lines, age depending on CEC) and observations (dots). Simulated time to reach 4 m depth (80–200 years, depending on CEC) is included in panel D (dashed line). Panels E and F show exchanger composition as equivalent fractions. (G–H) Simulated rates of $Al(OH)_3$ precipitation and $CO_2$ degassing to the unsaturated pore space after 100 years of liming *vs.* depth (negative values implies uptake by water). The groundwater table is near 5 m below surface.**

## Modelling downward front progression of the past

A reactive transport model, using PHREEQC-3 (*41*), was constructed to evaluate the progression of the alkalinity front from the time when liming began to its current position at 2 m depth, and its future continued downward movement. Inevitably, some uncertainty will be introduced by extrapolating current conditions to the past and future. However, such a model remains useful to validate the thermodynamic feasibility of the above reaction chain, quantify the processes involved, and to evaluate EW schemes on longer time spans.

The model was designed as transient, simulating the time needed for the alkalinity front to reach 2 m depth as a result of the percolation of water with the chemical composition observed above the alkalinity front, into an initially acidic top- and subsoil corresponding to the pre-liming situation. The model consisted of 50 cells, each representing a 0.1 m depth interval, and of which the upper 3 cells (0.3 m) represented topsoil. Water in each cell was assumed to be in equilibrium with the $P_{CO_2}$ (interpolated from observations) and Al-hydroxide, both as internal boundary conditions and fixed during the simulation, and a cation exchanger that was allowed to change in composition during the simulation. With CEC as an important determinant for the alkalinity front progression, two model runs were conducted, i.e., one with each set of top- and subsoil CECs (see above).

The comparison in Fig. 2A–F of summarized modelling results (lines) with observations (points) generally indicates reasonable correspondences, given that temporal variations in boundary conditions were not implemented in the model setup. The simulated front velocity range in the subsoil was 2.9 to 8.2 cm $yr^{-1}$, and thus in line with the results obtained by the retardation equation (Equation 1), indicating that the manual calculation addresses the major controlling processes controlling the front progression. Depending on the applied CEC, the simulated alkalinity front reached the observed 2 m depth after 55 to 130 years of simulation, in reasonable agreement with the site's liming history. In comparison, as illustrated by model lines in Fig. 2D, another 25–70 years will pass for the front to travel further 2 m downward, to 4 m depth (and 37–104 yr to the groundwater table 5 m depth; outside the scale of Fig. 2). The longer times required to travel the first 2 m are due to the ~10 times higher CEC of the topsoil (upper 0.3 m). Yet, the low subsoil CEC cannot be ignored because the subsoil is many times deeper than the topsoil. The front velocity range obtained by our assessment translates to 12–34 yr $m^{-1}$. Here, uncertainties in front velocity, in addition to those related to CEC, arise from assumptions of a stable percolation water chemistry (e.g., lines at 0 m depth in Fig. 2A–F), CEC (e.g., topsoil CECs are probably altered over time for example due to variations in soil organic carbon content or soil erosion) and recharge (which may also have varied over the last century (*42*)). Larger recharge rates, higher base cation concentrations in the infiltrating water, and lower CECs increase front velocities proportionally, and vice versa. Nevertheless, we propose that the full lines in Fig. 2AF reflects the situation many decades or a century into an EW scheme.

The model quantifies the rates and depth distributions of transformation of alkalinity to $CO_2$, its degassing to the pore space (Fig. 2H), and $Al(OH)_3$ precipitation (Fig. 2G). At the present alkalinity front position at 2 m depth, Al is displaced from the cation exchanger and precipitated as $Al(OH)_3$ at a rate amounting to 140 mmol $m^{-2}$ $yr^{-1}$ (sum of green columns in Fig. 2G). At the same depth, $CO_2$ is produced at a three times higher rate (420 mmol $m^{-2}$ $yr^{-1}$; sum of grey columns near 2 m depth in Fig. 2H) in response to the pH drop induced by $Al(OH)_3$ precipitation (cf. reaction 5). A lowering of pH from ~7 to 6.2 also takes place across the upper ~0.5 m of the profile due to uptake of $CO_2$ (900 mmol $m^{-2}$ $yr^{-1}$ down to 0.7 m depth; negative degassing in Fig. 2H) by infiltrating water as the water percolates to depths with gradually higher $P_{CO_2}$. This uptake solely forms carbonic acid, as lime dissolution is not explicitly integrated in the model (see above).

## Implications

The results from our temperate zone field site may be extended to other climatic zones. Topsoils affected by acidity account for 30% of the ice-free Earth land surface (*43*). 75% of these areas are affected by subsoil acidity, particularly prevalent in the humid tropics and subtropics. Even in developed countries subsoil acidity has been only partly ameliorated by agricultural liming (*43*).

A net uptake of atmospheric carbon via alkalinity generation will not commence before the acidity stored in both the topsoil and the unsaturated zone, below the topsoil, has been neutralized. Our study demonstrates that subsoil acidity may delay $CO_2$ sequestration via alkalinity formation past the short time scales necessary to help mitigate an imminent atmospheric $CO_2$ overshoot. This may apply even for soils which have had their topsoil acidity neutralized by long-term liming, as illustrated by our case. The 'subsoil delay', at our field site, was estimated to 12–34 years per meter of acidic subsoil, as depending on cation exchanger composition and capacity, and percolation water chemistry and flux. For an EW scheme planned on *acidic* topsoil, a 'subsoil delay' likewise must be assessed and added to the delay related to the topsoil's acidity neutralization. For an EW scheme planned on topsoil made *non-acidic* by preceding years of liming, many decades may still pass before the subsoil's (residual) acidity is neutralized. During the combined delay period, emissions from mining, comminution, transport and application (*44*) worsens the potential $CO_2$ overshoot. Before large-scale EW deployment at the cost of billions of USD (*7*), the effects of subsoil acidity therefore need to be carefully and directly evaluated. Society must demand that any carbon dioxide removal it funds is not merely assumed or promised for the future, but real, concurrent, and directly monitorable.

Another important implication of our study is that increases in topsoil alkalinity or electrical conductivity (*20*), or observed *in-situ* dissolution rates of amendment products, are poor proxies for actual increases in downward alkalinity fluxes as geochemical processes further down the flow path may return the alkalinity back into gaseous $CO_2$ or dissolved carbonic acid.

While the above results may be discouraging for the potential of EW, still agricultural soil management must play an essential role in climate mitigation as indicated by the increased weathering rates at the high $P_{CO_2}$ of soils (*11*, *25*) (Fig. 1A) and the associated large diffusive flux from the soil into the atmosphere. EW constitutes just one of numerous carbon farming practices (*45–47*) (e.g., no-tillage, biochar, crop rotation (*48*)). Future research should continue to address which existing carbon farming practice (cf. refs. (*46*, *47*, *49*, *50*)) is the best choice on acidic soil types, and whether EW on acidic soils enhances soil fertility over traditional liming, and potentially increases carbon sequestration by increasing soil organic carbon stocks (*17*).

## Materials and Methods

Four multi-level and multi-parameter sampling profiles (named A, B, C and D) were installed through a ~5 m thick sandy unsaturated zone underneath an agricultural field, 10 km south of the town of Ikast, Denmark. The profiles were centered on each of four experimental plots, comprising four 50 m × 50 m quadrants of a 100 m × 100 m area of the agricultural field. The profile locations were: A: 56°2'5.07"N, 9°10'24.40"E; B: 56°2'5.59"N, 9°10'21.66"E; C: 56°2'7.10"N, 9°10'22.68"E; D: 56°2'6.62"N, 9°10'25.31"E. The profile instrumentations were installed in parallel to the normal agricultural machinery operations (tilling, etc.) to allow for undisturbed soil treatment. Data shown in Figs. 1 and 2 are from profile C. For the duration of the experiment, barley (*Hordeum vulgare*) was grown in quadrant C.

### Pore water sampling

Each vertical profile of pore water was obtained by installing, in December 2010, ten PRENART Super Quartz PTFE suction cells, depth-distributed vertically over the unsaturated zone, and with a horizontal spacing of 0.5 m, parallel to the direction of ploughing. The suction cells were beforehand equipped with a septum at the upwards facing end, and mounted at the tip of Ø32 mm PVC tubes with the septum inside the tube. Installation holes (Ø50 mm) were made by percussion using a Geoprobe®. Before installation, 400 mL of a 2:1 wt:wt silica flour-water slurry was transferred to the bottom of the hole through a flexible tube and the suction cell-PVC tube assembly was then lowered into the slurry immediately after. This procedure ensured a good hydraulic connection for the suction cell to the pore water in the unsaturated zone sand. The small annulus around the PVC tube was afterwards filled with manufactured certified clean well pack quartz sand to 0.3 m below ground and then with topsoil to the surface. In each profile, pore water from the ten suction cells was collected ten times between May 2011 and March 2014. Pore water samples were collected by lowering a pre-evacuated (to 0.3–0.6 atm), acid-washed, long-and-thin 85 mL septum glass contained in a shuttle with the septum facing downwards into the well. At its lower end the shuttle held a spring-mechanisms with a double-ended needle. Upon unification between the shuttle and the septum of the suction cell, the needle ensured hydraulic contact between the unsaturated zone pore water and the evacuated inner of the septum glass. After 6–22 h the shuttle holding the septum glass was retrieved, and the septum glass was removed and stored at 5°C for transport. Following the fieldwork the water sample was transferred to 20 mL polyethylene vials through 0.20 µm cellulose acetate syringe filters (Sartorius Minisart) and stored at 5°C.

**Water table elevation**

Each profile was equipped with a monitoring well screened at ~6 m below the surface, installed similarly to the suction cells. Reference elevation of the well head was measured using a Trimble R8. Water table elevation was derived from water level tape measurements and pressure transducers equipped with loggers (DIVER®).

**Soil gas sampling**

Multilevel pore gas samplers consisted of bundles of OD 4 mm/ID 1.5 mm nylon tubing ranging in length from 0.11 to 5.88 m and equipped with ¼" sintered plastic Vyon® silencers. For installation, each bundle was transferred to an open-ended Ø43 mm (OD) polyethylene (PE) tube which was then lowered into a Ø50 mm installation hole, pre-made by percussion using a Geoprobe®. The PE tube was then retracted while the annulus was backfilled with quartz sand to 0.3 m below surface. The upper 0.3 m was backfilled using topsoil from the plow layer. Partial pressures of $CO_2$ and $O_2$ were measured in the field by connecting a GA2000 Gas Analyzer (Geotechnical Instruments, UK) directly to each nylon tube.

**Water content and soil temperature**

Soil moisture was estimated using 18 capacitance probes (Decagon 5TE sensors) installed May 30th–31st, 2011. The sensors were installed in quadrant A along a line parallel to the direction of ploughing (similar to the other equipment) with 0.20–1.10 m horizontal distance and at 14 different depths between 0.05–5.70 m below ground surface. Below 1 m depth, sensors were installed by drilling installation holes to a specific depth using the same percussion drill as used for the suction cells. The sensor was thereafter pushed into undisturbed sediment at the bottom of the drilled hole and carefully backfilled with material from the site. For samples above 1 m, the same procedure was followed, but the installation holes were either hand dug or hand augered. A continuous time series containing data at 20 min interval was collected from July 18th, 2011, till February 18th, 2015. Before this, only campaigns giving soil moisture profiles at three times were collected, i.e., May 31st, 2011, June 14th, 2011, and June 28th, 2011.

**Inorganic carbonate speciation**

Alkalinity was measured using a Gran-titration setup with a quantification limit of ca. 10 µeq $L^{-1}$ and a precision better than 2% and validated against distilled water and 1 meq $L^{-1}$ $NaHCO_3$ standards. The Gran-titration was performed using a Metrohm AG Herisau with a 10 mL burette. Total dissolved inorganic carbon (DIC) and pH were speciated in PHREEQC using measured pore water alkalinities, depth-interpolated values of values $P_{CO_2}$ obtained by pore gas measurements conducted on the day of water sampling, and temperature from the 5TE probes. Depth-interpolation was necessary as the pore gas samplers and pore water suction cups were not installed at the same depths.

**Analysis of major ions, Al and DOC**

Major cations and anions were measured by ion chromatography. Aluminum was measured using an Elan 6100DRC ICP-MS with PerkinElmer Elan-software (ver. 3) for data collection and quantification. Dissolved organic carbon (DOC) was analyzed on a Shimadzu TOC-VCN analyzer.

**Coring**

Sediment cores were collected in March 2014 from Profile C, using Ø5 polyethylene liners mounted in a stainless-steel tube pushed down by a jackhammer. Upon recovery, the cores were immediately sealed and stored at 5°C.

**Cation exchanger composition**

Cored sediments (see above) representing depth intervals 0.15–0.25, 0.55–0.65, 0.85–0.95, 1.2–1.3, 1.4–1.5, 1.57–1.67, 1.74–1.84, 1.9–2.0, 2.25–2.35, 2.6–2.7, 3.2–3.3 and 4.4–4.5 m in profile C were subjected to analysis for exchangeable cations using 1 M NaCl for exchangeable ammonium and 1 M $NH_4Cl$ for other cations. Extractant element concentrations incl. Ca, Mg, Na, K, Al, Fe and Mn were determined using an Elan 6100DRC ICP-MS; ammonium was determined spectrophotometrically. To enable correction for precipitation of released Al to Al-hydroxides, the pH of the $NH_4Cl$ extractant was recorded before and after extraction. A pH-increase was converted to an $OH^-$ loss that was divided by three to derive the equivalent potential Al loss by $Al(OH)_3$ precipitation. Also, extractant alkalinity changes were recorded to correct for Ca release from carbonate dissolution, and extractant $SO_4$ concentration increases were recorded to correct for Ca release by gypsum dissolution. The corrections obtained for Al-hydroxide precipitation, and for carbonate and gypsum dissolution, were in all cases insignificant. Extractant solutions were supersaturated for gibbsite but remained subsaturated for gypsum and for microcrystalline gibbsite, the latter evaluated as 'Al(OH)3(a)' defined in the *wateq4f.dat* PHREEQC database file.

**Sediment DOC extractions**

Adsorbed concentrations of DOC to sediment samples from depths 1.10–1.20, 1.30–1.40, 1.50–1.57, 1.67–1.74, 1.84–1.90, 2.00–2.08, 2.08–2.17, 2.17–2.25, 2.35–2.42, 2.51–2.60, 2.70–2.80 and 2.90–3.00 m in profile C were extracted by 24 h shaking of 10 g of non-dried sediment in 30 mL of a 0.01 M NaOH solution. Gravimetrical water content was determined on parallel sediment samples. Following filtration using 0.20 µm EXP PES Minisart syringe filters, the DOC concentrations of the extracts were analyzed on a Shimadzu TOC-VCN analyzer.

**Reactive transport modeling**

Additional details of the reactive model described in the main text follow below. The *wateq4f.dat* database was used. The stability constant for the Al-hydroxide phase applied with EQUILIBIRUM_PHASES in all cells was $K_{\text{Al-hydroxide}} = 10^{-33.7}$, as defined for the reaction $Al(OH)_3 \leftrightarrow Al^{3+} + 3OH^-$. The $P_{CO_2}$ applied in each cell was interpolated from the observed partial $CO_2$ pressures (Fig. 1A), resulting in a gradual increase down to 65 cm depth where after it was kept constant with depth, in accordance with the above analytically assessed insignificant effect of $CO_2$ generation at the alkalinity front. Atmospheric $P_{CO_2} = 10^{-3.4}$ atm (400 ppmv) was assumed at the surface (depth 0 cm). $P_{CO_2}$ was defined with EQUILIBRIUM_PHASES; the following values were used: $P_{CO_2} = 10^{-2.59}$ atm in cell 1 (cell mid-point 5 cm depth); $P_{CO_2} = 10^{-2.07}$ atm in cell 2 (cell mid-point 15 cm depth, etc.); $P_{CO_2} = 10^{-1.77}$ atm in cell 3; $P_{CO_2} = 10^{-1.60}$ atm in cell 4; $P_{CO_2} = 10^{-1.50}$ atm in cell 5; $P_{CO_2} = 10^{-1.48}$ atm in cell 6; $P_{CO_2} = 10^{-1.47}$ atm in cells 7–50. As initial conditions, the observed water chemistry in the current acidic zone below the alkalinity front, and an initial cation exchanger composition in equilibrium with this water chemistry, were used. As the upper boundary condition, percolation water with an average chemical composition matching that observed above the alkalinity front was added, at a fixed percolation rate producing a 1-year water residence time in the modelled 5 m profile, cf. the 500 mm recharge to a soil column with a 0.1 volumetric water content mentioned in the main text. One common set of cation exchange coefficients and one common stability constant for the Al-hydroxide were fitted by trial-and-error. Table 1 shows the common set of cation exchanger coefficients applied in all cells. The commented input file for PHREEQC is included in the Supplementary Materials.

**Table 1: Common set of cation exchanger coefficients used in the reactive transport model (trial-and-error fitting).**

| Reaction defined in PHREEQC | log *K* |
|---|---|
| $Na^+ + X^- \leftrightarrow NaX$ | 0.0 |
| $K^+ + X^- \leftrightarrow KX$ | 1.15 |
| $Al^{3+} + 3X^- \leftrightarrow AlX_3$ | 0.15 |
| $H^+ + X^- \leftrightarrow HX$ | 0.45 |
| $Ca^{2+} + 2X^- \leftrightarrow CaX_2$ | –1.1 |
| $Mg^{2+} + 2X^- \leftrightarrow MgX_2$ | –1.08 |

**Data treatment and plotting software**

Data files were established in Microsoft Excel and treated with the Python pandas package using the Spyder GUI. Figures were made using packages seaborn and matplotlib.pyplot. Boxplots were made with seaborn. The central box begins and ends at the first and third quartile values, respectively, of the distribution. The central line represents the median. The whiskers extend to the furthest data points within 1.5 times the interquartile range from the quartiles; points lying beyond the whiskers are plotted individually. When plotting pore gas $CO_2$ (Fig. 1A) and carbonate speciation (Fig. 1B) the seasons named 'summer', 'autumn', 'winter' and 'spring' were defined to cover day-of-year intervals [150;240[, [240;310[, [310;60[ and [60;150[, respectively. This corresponds roughly to the periods June to August as summer, September to primo November as autumn, primo November to February as winter, and March to May as spring. Fig. 1A contains data from 37 pore gas samplings conducted between February 2011 and December 2012. Seasonality in Fig. 1A was further illustrated drawing curves with data from 19th June (i.e., summer), 25th October (autumn), 19th December (winter) and 28th March (spring) of 2012. The carbonate speciation illustrated in Fig. 1B is for samples collected on 19th September (autumn) and 13th March (spring) of 2012.

## Acknowledgements

For technical and academic assistance, guidance and discussions we thank Pernille Stockmarr, Per Jensen, Christina R. Lynge, Jens Bisgaard, Johan Mølby, Erik Rønn Lange, Bent Hansen, Flemming Larsen, Claus Beier, Eike Marie Thaysen, Iver Jakobsen, Mariana Garcia Iwasaki, Tom Christensen and Tobias B. Uglebjerg. Funded by Danish Strategic Research Council (DSF-09-067234). All data needed to evaluate the conclusions in the paper are present in the paper and/or the Supplementary Materials.

## Author contributions

SJ installed water and gas collection systems, and collected samples under supervision of DP, RJ and PA. ML installed soil moisture sensors, and managed collection and curation of their data. PA supervised supporting gas sampling and conducted in-house gas analysis. SJ, DP and RJ conceived the story of the paper. SJ curated data, prepared figures, and wrote the first draft. All authors interpreted results and revised the manuscript.

# Supplementary Materials for

## Subsoil acidity causes long delays in inorganic carbon sequestration by Enhanced Weathering

Søren Jessen* *et al.*

*Corresponding author: Email: sj@ign.ku.dk, Phone: +45 5137 0693

**This PDF file includes:**

Figs. S1 to S5

Supplementary Texts 1 and 2

- Calculation of added partial $CO_2$ pressure due to alkalinity neutralization at 2 m depth, incl. Fig. S6.
- Downward progression velocity of alkalinity front.

References (51 to 56)

**Other Supplementary Materials for this manuscript include the following:**

Data S1: Data availability statement and overview

**Fig. S1**

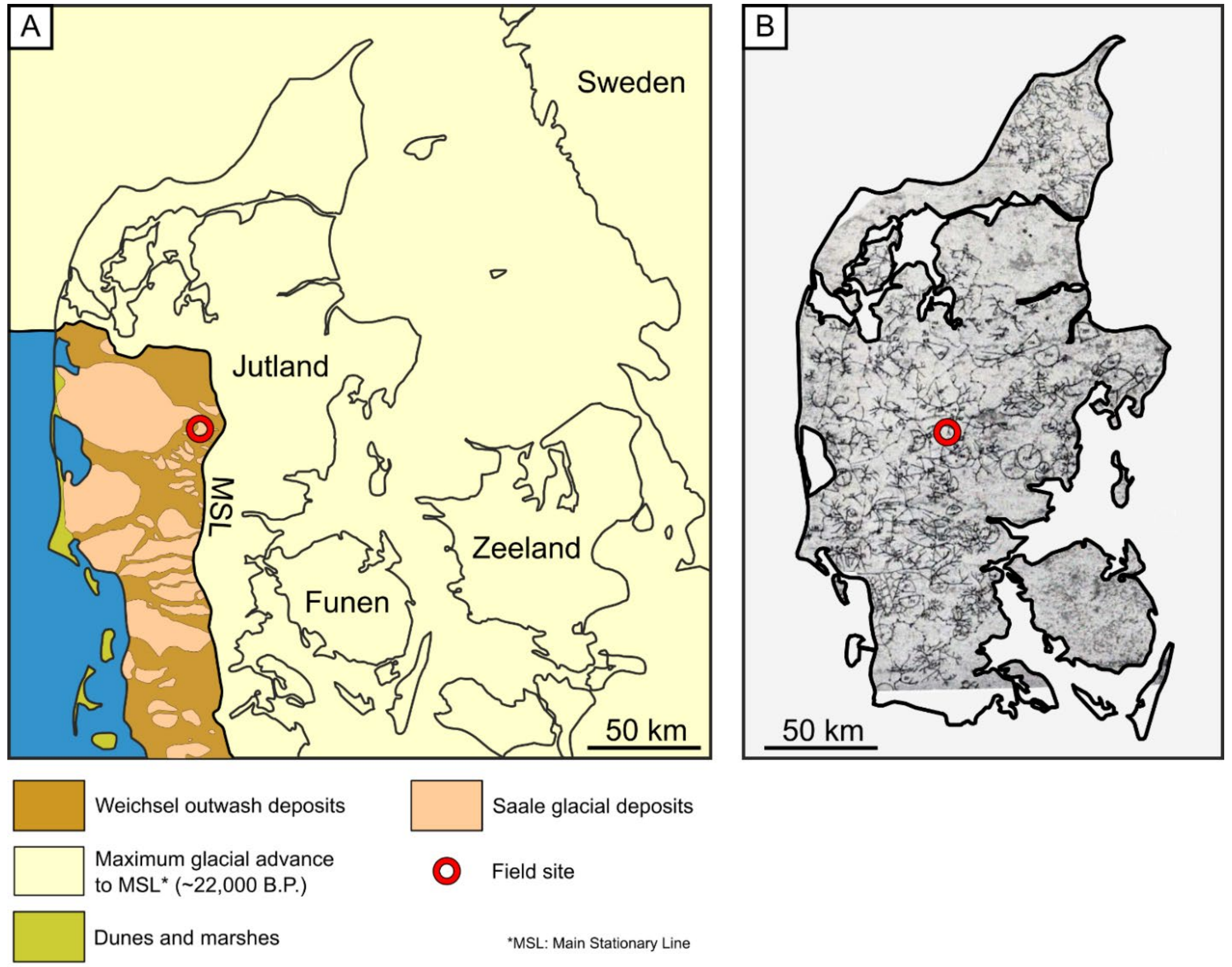

Fig. S1. (A) The location of the field site relative to Weichsel and Saale deposits. The field site lies within an area of outwash sand of Saale. Accordingly, the field site lies in sandy sediments exposed to weathering for some tens of thousands of years longer than the more recent outwash sands of Weichsel. (B) Hand-drawn map of marl train tracks in Jutland (modified from *51*), laid out by more than 100 independent marl companies. Each company typically existed for just a few years. Most of the tracks were temporary narrow-gauge rails, often made of rented materials, which were moved to another location upon closure of a track (*51*). The marl tracks and the social shared workload by all marl receivers along a track enabled also small farms to treat their soils with marl (*52*). The first two companies respectively created the Damholt-Hodsager marl track and the Grindsted-Grene marl track; these tracks were first used in 1879 and 1886, respectively (*51*). These tracks, both 20 km long, began at marl pits located 15 km north and 30 km south of the field site, indicating that the field site was near the first marl tracks. In the period that followed, the marl track network indicated in Fig. SI1B grew, and then dwindled again—the last marl track was abandoned by the end of the 1930'ies. At that time trucks had taken over most of the marl and lime transport, and lime had become the preferred amendment product (*52–54*).

**Fig. S2**

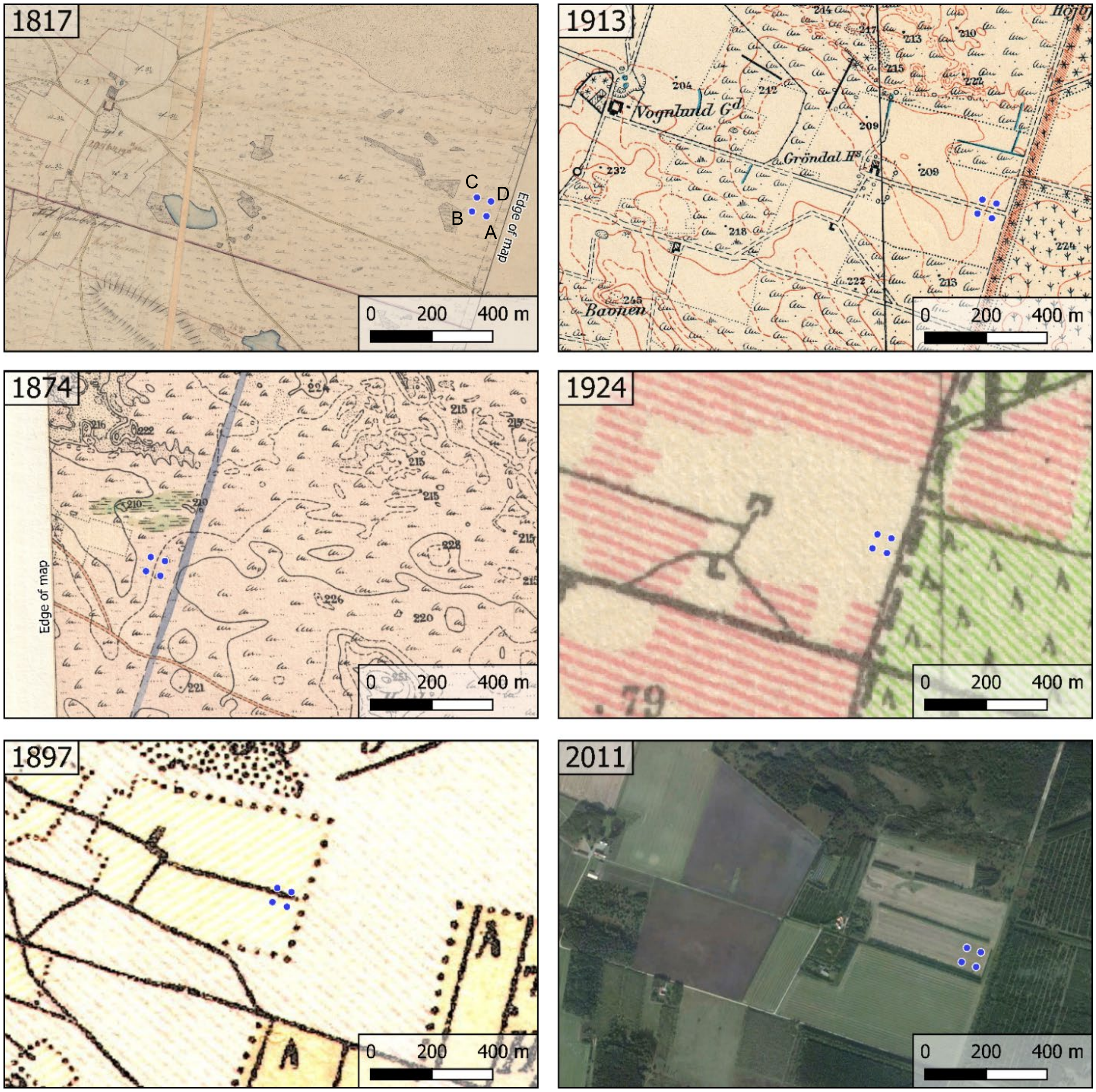

Fig. S2. The location of the four profiles (A, B, C and D; blue dots) superimposed on historical maps from 1817 to 1924, and on an orthophoto from 2011 when the profiles were recently installed (north up). In 1817, the four profiles were within an area of heathland with heather vegetation (heather-symbol: ) and several hundred meters from the nearest cultivated land. In 1874, agricultural fields (marked with dotted lines) were within a distance of 160 m from profile C. In 1897 all four profiles were within an agricultural area. The maps hence indicate that cultivation of the field site began in the late part of the 19$^{th}$ century. In support, the systematic cultivation of the large heath areas of western Jutland was the specific aim of The Danish Heath Company, founded in 1866. The Danish Heath Company supported the effort to amend the reclaimed heath areas with marl and lime (cf., Figs. S1B and S3).

**Fig. S3**

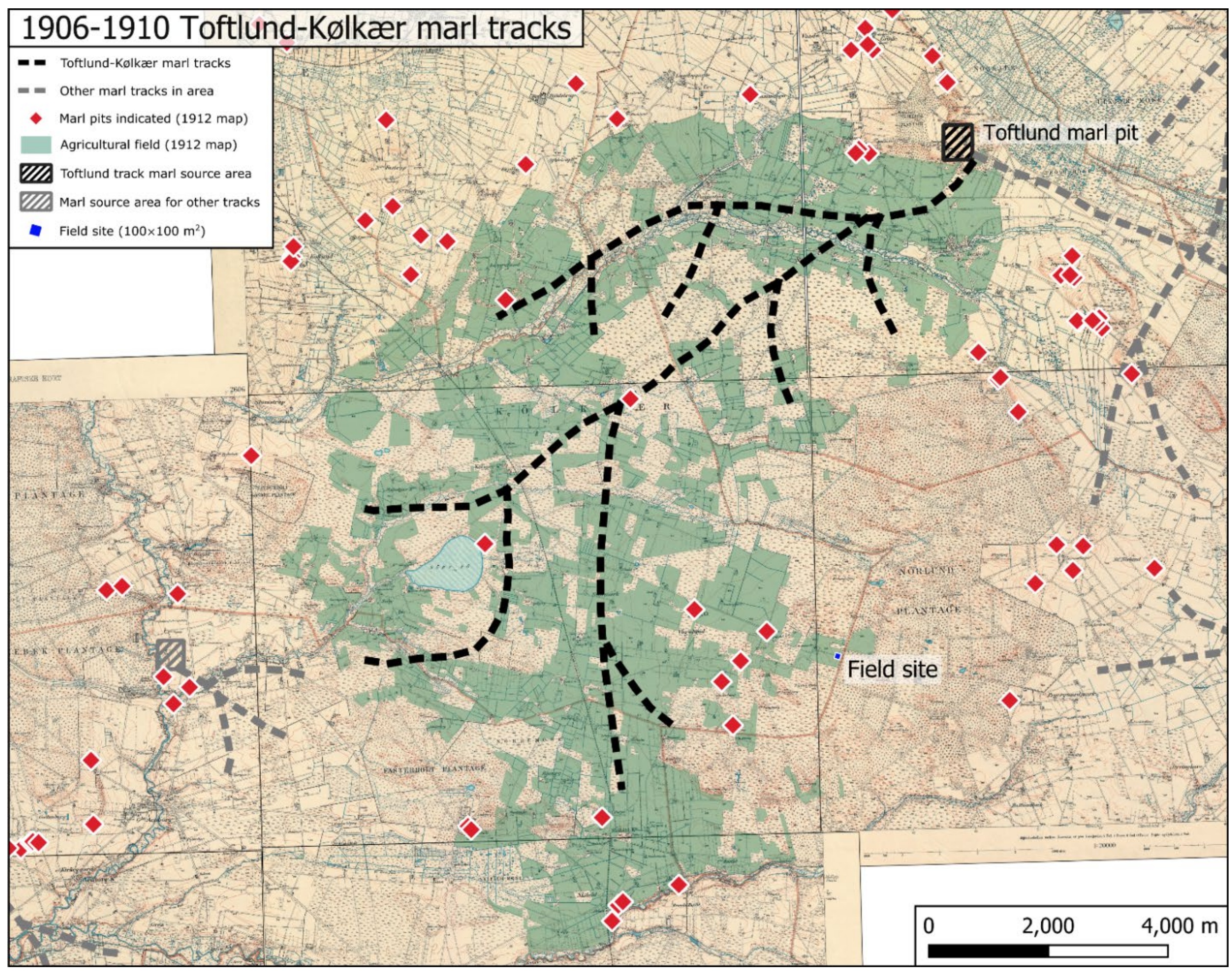

Fig. S3. Location of the Toftlund-Kølkær marl tracks (black dashed lines) and other tracks (grey dashed lines) in the area, as superimposed from hand drawings by Bent Hansen (*pers. comm.*) onto maps from 1912. Out of more than 100 companies created to distribute marl in the late 1800's and early 1900's, the Toftlund-Kølkær marl track company was formed as the sixth (Fig. SI3). This company existed between 1906 and 1910 and created the Toftlund-Kølkær marl tracks that brought out 52000 $m^3$ before it closed in 1910 (*54*). The green shaded agricultural areas in the 1912-maps are areas that we evaluate as being potential target areas for the marl, based on vicinity to the Toftlund-Kølkær marl tracks and relative absence from other sources (marls tracks or pits). The area of the green shaded land is 5300 ha. Thus, as a rough estimate, each hectare of reclaimed heath potentially in average received up to 9.8 $m^3$ of marl via marl tracks. In comparison, the Annual Report by The Danish Heath Company in 1939 stated that a total of 21.000.000 $m^3$ marl had been brought out to an estimated area of 200 square miles (1 square mile = 5681 ha), yielding an average of 18.5 $m^3$/ha (*55*). These numbers indicate that a strong and effective effort was made to increase soil pH of reclaimed heath.

**Fig. S4**

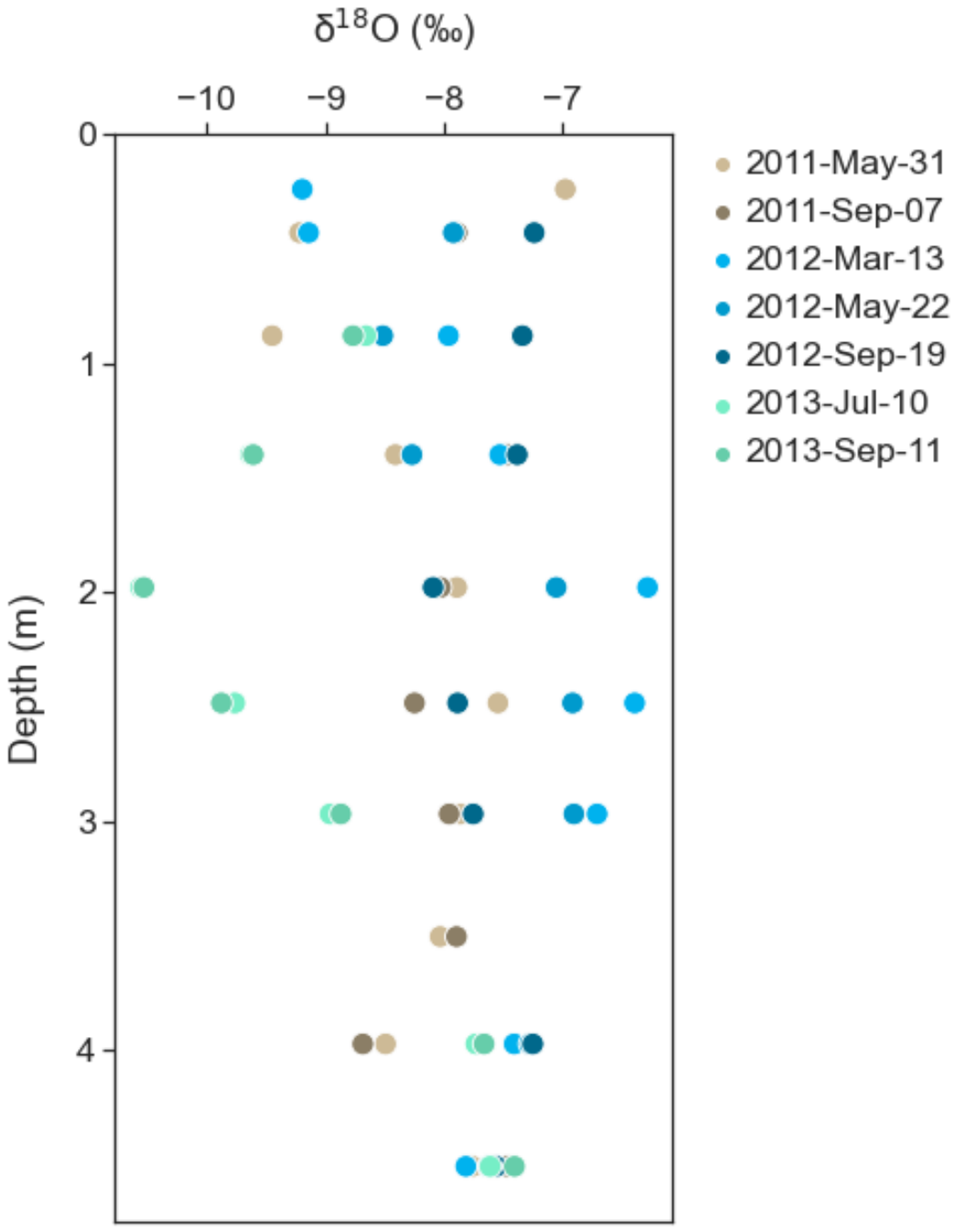

Fig. S4. Variation in $\delta^{18}O$ of pore water vs. depth in profile C over the course of approximately two years. At each sampling depth, pore water $\delta^{18}O$ remained relatively stable over summer periods, between May and September, when percolation is usually low, whereas larger shifts in isotopic values occurred over the winter periods, when percolation normally takes place.

**Fig. S5**

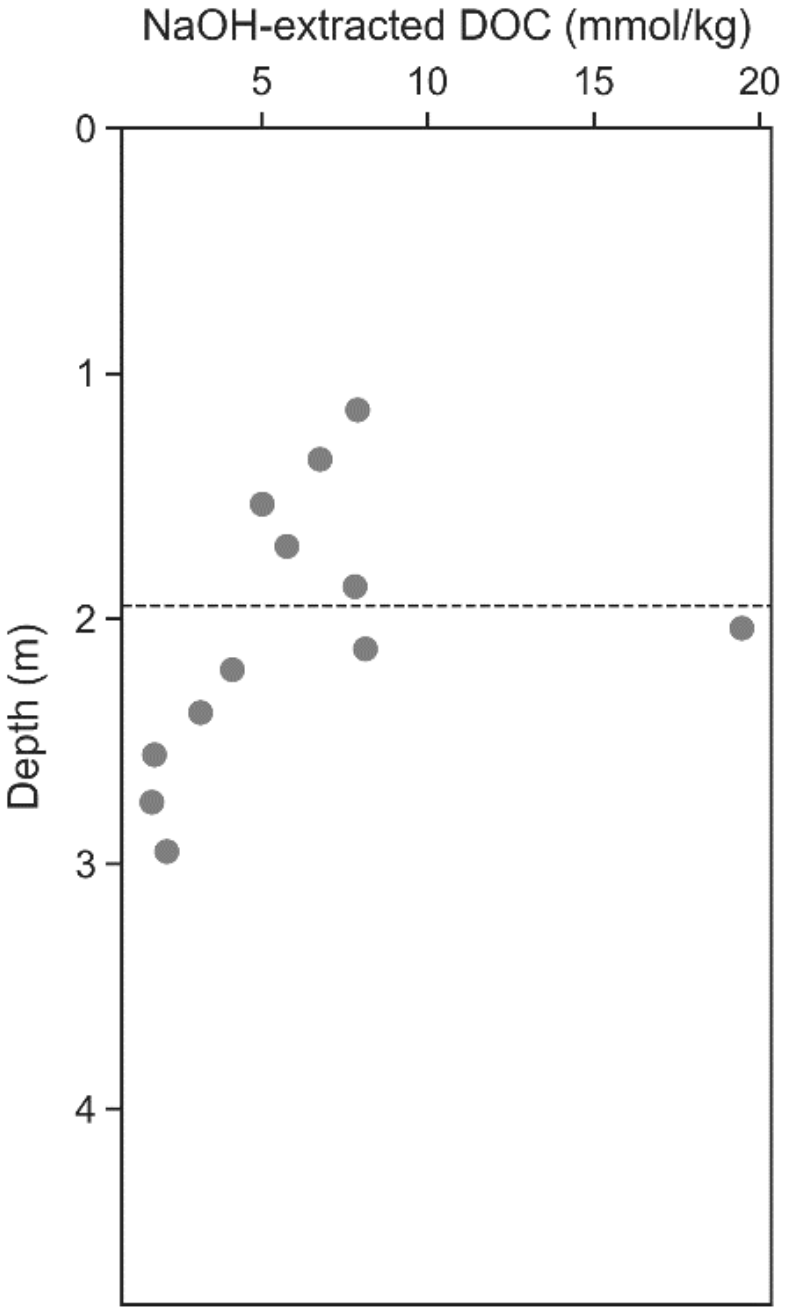

Fig. S5. NaOH-extractable DOC *vs.* depth in profile C. Dashed line indicates the depth of the alkalinity front.

**Supplementary Text 1**

Calculation of added partial $CO_2$ pressure due to alkalinity neutralization at 2 m depth

It can be shown that the added partial pressure due to degassing of $CO_2$ produced by the combined cation exchange and Al-hydroxide equilibration is insignificant. Fick's first law:

$$F = D_e \times \frac{dc}{dx} \Leftrightarrow dc = \frac{F}{D_e} \times dx$$

allows calculation of the change in concentration *dc*, for a given flux *F*, distance *dx*, and effective diffusion coefficient $D_e$.

In our case, the flux *F* of 500 mmol/m$^2$/yr (=1.59×10$^{-8}$ mol m$^{-2}$ s$^{-1}$) is generated a 2 m depth. We assume that this $CO_2$ diffuses upwards towards a steady state boundary condition at a depth of 0.5 m where prevailing soil conditions keep an arbitrary partial pressure of $CO_2$.

Using the lower end of the measured effective gaseous diffusion coefficient range 1.2–6×10$^{-6}$ m$^2$ s$^{-1}$ (*1*), we calculate the steady state *dc* at 2 m depth:

$$dc = \frac{1.59 \times 10^{-8} \frac{mol}{m^2 s}}{1.2 \times 10^{-6} \frac{\mathrm{m}^2}{\mathrm{s}}} \times (2\,m - 0.5\,m) = 0.020\,mol/m^3$$

To relate this value to the pressure of the atmosphere, we first calculate the number of moles of gas molecules *n* in a unit volume *V* of 1 m$^3$ at an atmospheric pressure *P* of 101300 Pa and at 10°C using the ideal gas law:

$$\frac{n}{V} = \frac{P}{RT} = \frac{101300\,Pa/atm}{8.314\,J/K\,mol \times 283.15\,K} = 43.03\,mol/atm\,m^3$$

In units of atm *dc* then becomes:

$$dc = \frac{0.020\,mol/m^3}{43.03\,mol/atm\,m^3} = 4.6 \times 10^{-4}\,atm$$

This *dc* is too small to be observable in most subsoil $P_{CO_2}$ profiles with, typically, 10–100 times higher $P_{CO_2}$ (e.g., Fig. 1A of the main text), and even less observable when considering the dynamic and transient nature of real profiles (cf. Fig. 1A in the main text). The *dc* would be even smaller if the higher end (i.e., 6×10$^{-6}$ m$^2$ s$^{-1}$) of the observed $D_e$ range had been applied in the calculation.

The red line in Fig. S6 illustrates how the $CO_2$ concentration *dc* in units of mol m$^{-3}$ would increase from 0 to 0.020 mol m$^{-3}$ as the depth increases from an arbitrary $P_{CO_2}$ at the fixed boundary condition at 0.5 m depth to the depth of the alkalinity transformation (i.e., $CO_2$ production) at 2 m. The green line in Fig. S6 illustrates the effect that this *dc* would have on $CO_2$ profile assuming a $P_{CO_2}$ of 0.02 atm below 0.5 m depth without carbonate alkalinity transformation to $CO_2$ at 2 m depth. Clearly, no peak in $P_{CO_2}$ would be observable at 2 m depth.

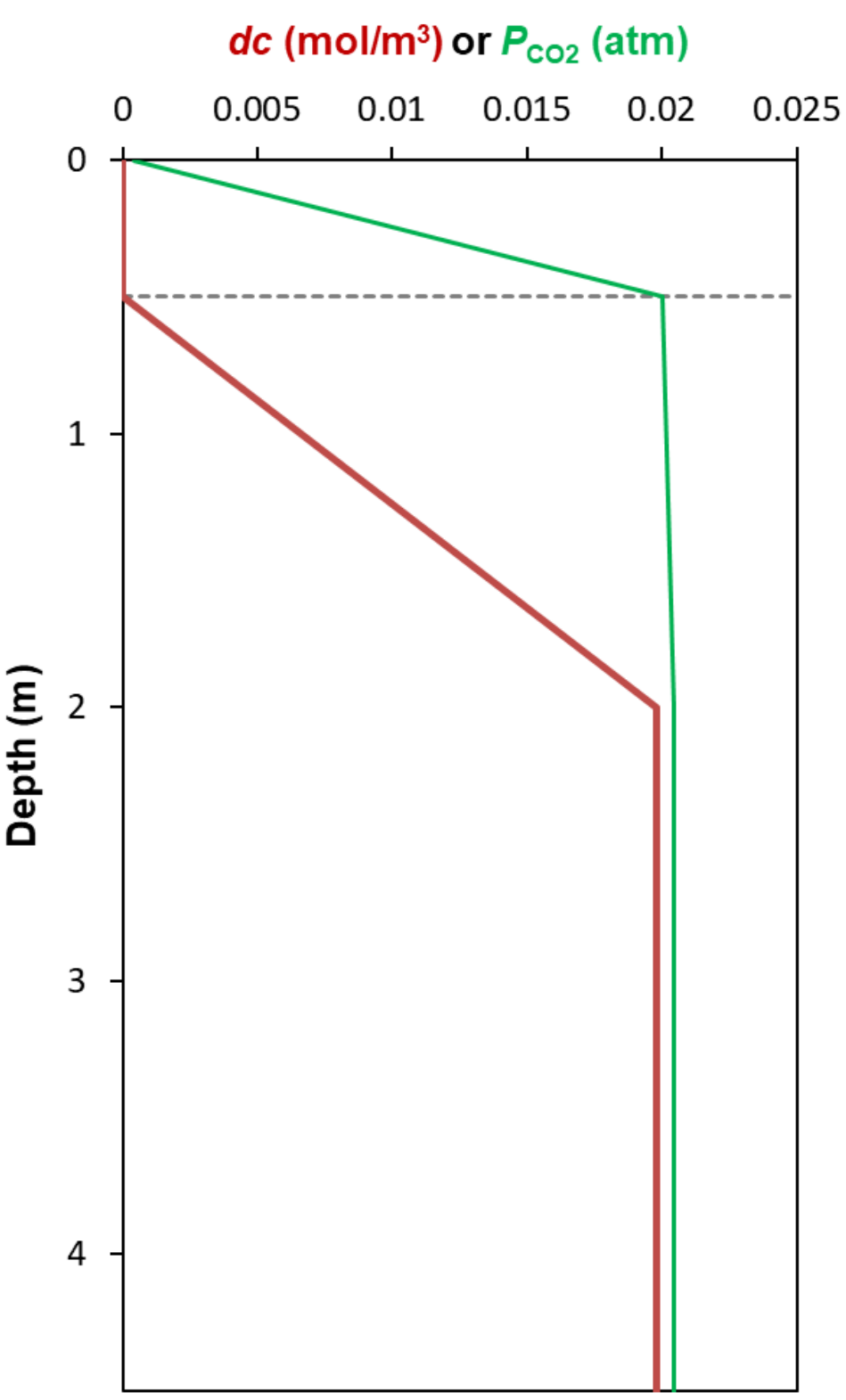

Fig. S6. Illustration of the effect of $CO_2$ production by alkalinity transformation *vs.* depth in profile C, as calculated by Fick's first law. The red line shows the *dc* needed to support an upwards $CO_2$ flux of 500 mmol $m^{-2}$ $yr^{-1}$, using an effective gas diffusion coefficient of $1.2\times10^{-6}$ $m^2$ $s^{-1}$. The green line illustrates effect of the $CO_2$ production at 2 m depth on the shape of a $P_{CO_2}$ profile. The diffusion from 0.5 m depth to the surface is not calculated but drawn by connecting with a straight line to the atmospheric $P_{CO_2}$ at the surface.

**Supplementary Text 2**

Downward progression velocity of alkalinity front

The current downward progression of the alkalinity front can be derived from the mass transfer between the exchanger-bound Ca and Mg and the solute Ca and Mg concentrations, using the formula (*56*):

$$R_{\text{front}} = 1 + \frac{dq}{dc}$$

where $R_{\text{front}}$ is the retardation coefficient and $dq$ and $dc$ are the changes in, respectively, exchanger-bound and solute Ca and Mg concentrations, both in units of mmol per L of contacting porewater. The concentrations used to calculate $dc$ equals those of the reactive transport model results (above *vs.* below the front) which correspond fairly well to the observed dissolved concentrations, c.f. Fig. 2A and B in the main text. For $dc$ we have:

$dc$ = dissolved sum of Ca and Mg above minus below the front
= (0.6 – 0.25) + (0.15 – 0.09) mM = 0.40 mM = 0.80 meq/L

For $dq$ we have:

$dq$ = exchanger-bound Ca and Mg above minus below the front
$= (\beta_{\text{Ca+Mg, above front}} - \beta_{\text{Ca+Mg, below front}}) \times \text{CEC (meq/100 g)} \times (1-\varepsilon) \times \rho_{\text{grain}} / \varepsilon_w$

For the sand beneath the topsoil, a volumetric water content $\varepsilon_w$ of 0.1 L pore water per $dm^3$ bulk sediment, and porosity $\varepsilon$ of ca. 0.3 was observed. The grain density $\rho_{\text{grain}}$ is assumed equal to that of quartz, i.e., 2.65 $kg/dm^3$. The observed changes, across the front, in $\beta_{Ca}$ and $\beta_{Mg}$ was evaluated from Fig. 2E in the main text: $\beta_{Ca}$ drops from ca. 0.55 to ca. 0.2 across the front, while $\beta_{Mg}$ drops from ca. 0.12 to ca. 0.03. Considering the span in measured subsoil avg. CEC of 0.56 (from *sum* of adsorbed cations) to 1.6 meq/100 g (*direct* Na displacement; see main text) we calculate two corresponding values of $dq$:

$$dq_{\text{sum}} = (0.55-0.2 + 0.12-0.03) \times 0.56 \text{ meq/100 g} \times (1-0.3) \times 2.65 \text{ kg/dm}^3 \times 10 \text{ hg/kg} / 0.1 \text{ L/dm}^3$$

$$dq_{\text{sum}} = 45.4 \text{ meq/L}$$

$$dq_{\text{direct}} = (0.55-0.2 + 0.12-0.03) \times 1.6 \text{ meq/100 g} \times (1-0.3) \times 2.65 \text{ kg/dm}^3 \times 10 \text{ hg/kg} / 0.1 \text{ L/dm}^3$$

$$dq_{\text{direct}} = 131.3 \text{ meq/L}$$

which yields retardation factors of:

$$R_{\text{front}}^{\text{sum}} = 1 + \frac{0.80\ meq/L}{45.4\ meq/L} = 58 \quad \text{and} \quad R_{\text{front}}^{\text{dir}} = 1 + \frac{0.80\ meq/L}{131.3\ meq/L} = 165$$

Finally, using the formula $v_{\mathrm{front}} = v_{\mathrm{H_2O}}/R_{\mathrm{front}}$, with a pore water velocity of 5 m/yr, the front velocity range becomes 3.0 to 8.7 cm/yr. Using values for dissolved and exchanger-bound Ca alone in the above calculation results in slightly lower retardation factors of 54 and 155, and slightly higher velocities of 3.2 to 9.2 cm/yr. For comparison, the simulated front velocity range in the subsoil was 2.9 to 8.2 cm/yr.

## References 51 to 56

**Data**

Data S1: Data availability statement

At Copenhagen University's Harvard Dataverse:

https://dataverse.harvard.edu/previewurl.xhtml?token=b06af002-e921-472a-bdca-4fe8c0ac19bf

tables with observational data from all profiles, figures, and the PHREEQC-3 model file are available in the following folder tree structure:

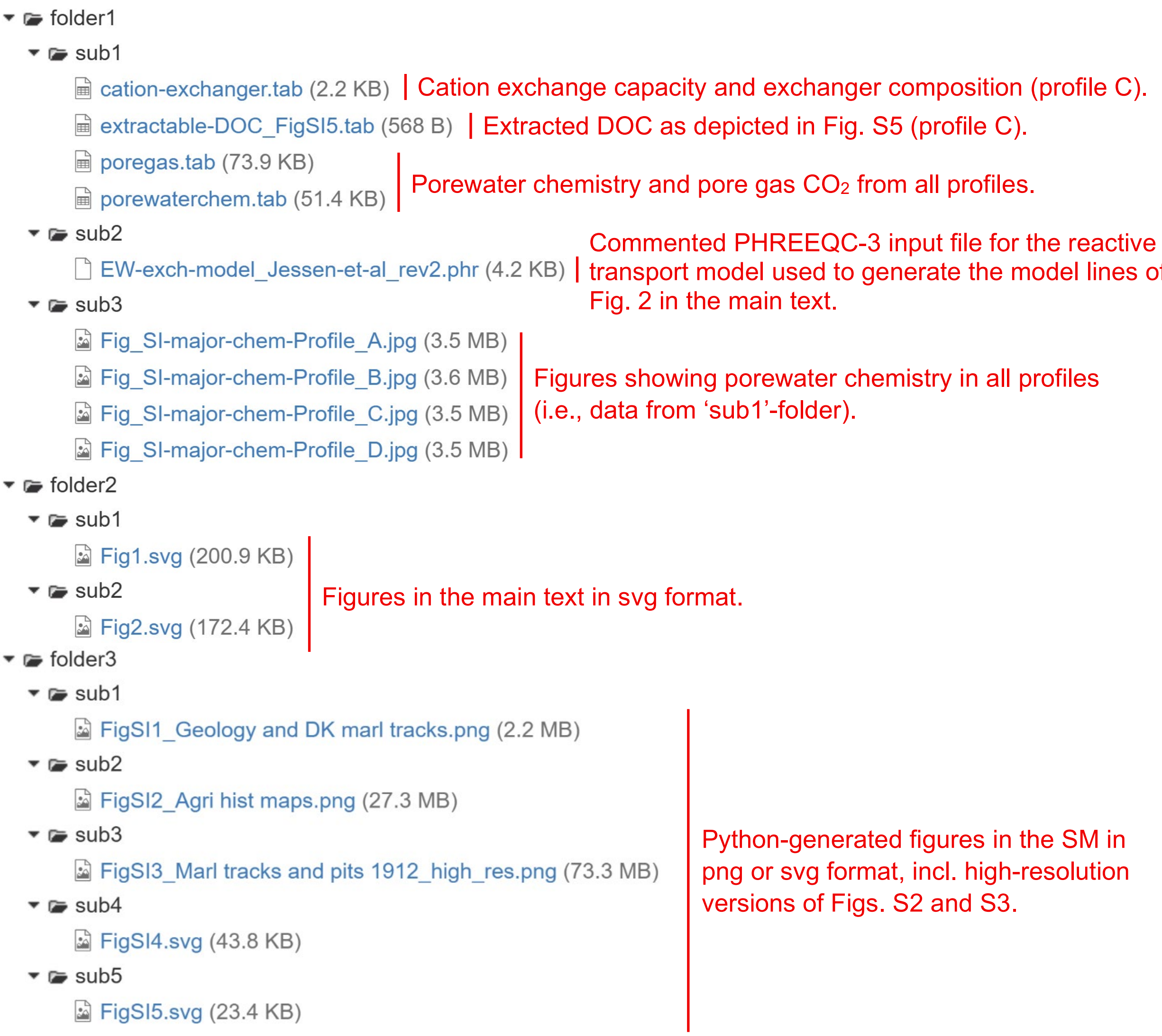